\documentclass[twocolumn,preprintnumbers,amsmath,amsfonts,amssymb,notitlepage,showpacs,pra]{revtex4-2}
\usepackage{color}

\date{\today}
\usepackage{graphicx}

\usepackage{hyperref}

\def \be{\begin{equation}}
\def \ee{\end{equation}}
\def \ba{\begin{array}}
\def \ea{\end{array}}
\def \bea{\begin{eqnarray}}
\def \eea{\end{eqnarray}}

\begin{document}
\title{
Rotating Bose gas dynamically entering the lowest Landau level
}
\author{Vaibhav Sharma}
\email{vs492@cornell.edu}
\author{Erich J Mueller}
\email{em256@cornell.edu}
\affiliation{Laboratory of Atomic and Solid State Physics, Cornell University, Ithaca, New York}
\begin{abstract}

Motivated by recent experiments, we model the dynamics of a condensed Bose gas in a rotating anisotropic trap, where the equations of motion are analogous to those of charged particles in a magnetic field. As the rotation rate is ramped from zero to the trapping frequency, the condensate stretches along one direction and is squeezed along another, becoming long and thin. When the trap anisotropy is slowly switched off on a particular timescale, the condensate is left in the lowest Landau level. We use a time dependent variational approach to quantify these dynamics
and give intuitive arguments about the structure of the condensate wavefunction. This preparation of a lowest Landau level condensate can be an important first step in realizing bosonic analogs of quantum Hall states.

\end{abstract}

\maketitle

\section{Introduction}\label{intro}

For many years the 
ultracold atom community has been attempting to
use neutral atoms to explore 
the physics  of  2D electron gases in 
magnetic fields.
The most exciting possibility is  
realizing fractional quantum Hall physics  and 
observing non-abelian phases of matter~\cite{cooperreview}. A recent experiment at MIT~\cite{fletcher2019geometric} made an important step towards achieving this goal.
They found that when
rotating a
Bose gas in an anisotropic trap,
the resulting non-equilibrium dynamics
can drive the atoms into the lowest Landau level, a pre-requisite for observing the quantum Hall effect.
Motivated by this observation, we model the dynamics of a rotating Bose gas in an anisotropic trap using a time dependent variational wavefunction. 
We characterize the protocols which drive the atoms into the lowest Landau level.

There are a number of techniques for using neutral atoms to emulate the properties of charged particles in a magnetic field:  These
include optical dressing~\cite{dressedlight}, shaking~\cite{shakinglattice} and rotating~\cite{fetter}. 
Here we focus on rotation, where the connection with electronic systems
 can  be understood classically: In the rotating frame, the Coriolis force, $F_{\rm c}=2m\,\vec{v}\times \vec{\Omega}$ has the same form as the Lorentz force $F_{\rm L}=q\, \vec{v}\times \vec{B}$ on a charged particle $q$ in a magnetic field $\vec{B}$.  Here $\vec{\Omega}$ is a vector whose magnitude is the rotation frequency, and direction is the axis of rotation (which we take to be $\hat z$).
 One can formally identify $q\vec{B}=2 m\,\vec{\Omega}$.  The rotating system also experiences an outward centrifugal force, which reduces the strength of the in-plane harmonic trapping potential.  If the trapping frequency equals the rotation frequency, $\omega=\Omega$, the effective trapping frequency vanishes, and in the rotating frame the neutral atoms have equations of motion which are identical to electrons in a uniform magnetic field.  The quantum mechanical eigenstates correspond to highly degenerate Landau levels, whose energies are separated by 
$2\hbar\Omega$. 
The resulting behavior is quite rich; for example, rotating superfluids display vortex lattices similar to those of
type-2 superconductors in magnetic fields~\cite{vortexlattice,vortexlattice1,vortexlattice2} and also exhibit a rotational Meissner effect~\cite{heliummeissner}.

It is meaningless to talk of the rotation frequency of a rotationally symmetric potential.  Thus, 
following
the experimental approach in Ref.~\cite{fletcher2019geometric}, we include a small rotating elliptical deformation which stirs the cloud.  In the rotating frame this potential takes the form  $V_\varepsilon=-\varepsilon m \omega^2 x y$,
and can be described in terms of
an electric field $q\vec{E}=-\vec{\nabla} V_\varepsilon$. In the steady state, 
the bosons will experience a Hall drift with a velocity proportional to $\vec{E}\times\vec{B}$, like classical charged particles in crossed electric and magnetic fields. This drift ``squeezes" the gas into a long thin shape, as shown in Fig.~\ref{densityprofile}. 
Remarkably, one can  interpret the atomic state as being confined to a distorted Landau Level.  

If the trap deformation is turned off at the appropriate time, the wavefunction of each atom ends up in the traditional lowest Landau level.  
The turn-off must occur adiabatically on a timescale which is moderately larger than $1/(\varepsilon\omega)$, which  is the characteristic dynamical time. 
If the turn-off is too slow, however, non-linear terms drive the system back out of the lowest Landau level.
   This can also be interpreted in terms of 
   transitions between levels which cross during the dynamics,
   and requires that the deformation is turned off on a timescale which is small compared to $8/(\varepsilon \omega) \ln(8/\varepsilon)$.

It is worth noting that the number of single-particle states in the Landau level is proportional to the area of the system, and it represents a large phase space.  One cannot equate being in the lowest Landau level with being in a quantum Hall state.  Nonetheless, driving the system into the lowest Landau level is an important first step in realizing a cold atom quantum Hall effect.  Furthermore, the dynamics in the lowest Landau level are quite rich, and are a useful area of study on their own~\cite{shlyapnikov,bosonsinlll}.


There have been numerous other works exploring rotating Bose gases in order to access the quantum Hall regime \cite{cooperreview}. The lowest Landau level was nearly achieved by Schweikhard \textit{et al}.~\cite{rotatingbosegaslowestlandau} by using an evaporation technique.
In the few-body context,
Gemelke \textit{et al}.~\cite{rotatingfqhe} 
found a driving protocol which not only produced lowest-Landau level wavefunctions, but even analogs of the Laughlin
state. 
Stirring protocols similar to Ref.~\cite{fletcher2019geometric} have been previously used, but the lack of in-situ imaging 
meant they could not study the physics explored here
\cite{Dalibard1,Dalibard2}.


Our paper is organized as follows. In Sec.~\ref{model}, we introduce the model. In Sec.~\ref{wavefnansatz}, we write down our time dependent wavefunction ansatz to model the dynamics. In Sec.~\ref{wavefndynamics}, we use the time dependent variational principle to derive the equations of motion for the wavefunction parameters. In Sec.~\ref{dynamicssection}, we solve for the dynamics of the system and interpret the results.

\section{Model}\label{model}

We consider bosons at zero temperature in a harmonic trap with frequencies $(\omega_x,\omega_y,\omega_z)$. For $ \omega_x,\omega_y = \omega$ and $\omega_z\gg \omega$, the z-motion is frozen and one  effectively has a two dimensional Bose gas.

We add a rotating deformation potential given by $V_r(t)=-\varepsilon m\omega^2  X(t)Y(t)$ where, $X(t)= x\cos{\phi(t)}+y\sin{\phi(t)}$ and $Y(t)= -x\sin{\phi(t)}+y\cos{\phi(t)}$. We assume that the rotation rate $\Omega(t)=d\phi/dt$ is slowly varying.  The strength of the deformation is given by the dimensionless variable  $\varepsilon$. Note, if $\phi$ is phase shifted by $\pi/4$, then $V_r(t)\to 
\varepsilon m\omega^2 (X(t)^2-Y(t)^2)/2$, which is a form used in other papers~\cite{fletcher2019geometric,shlyapnikov}.

In the mean field regime, the condensate wavefunction $\psi$ obeys the 2D Gross-Pitaevskii equation (GPE), 
\begin{equation}
    i\hbar\frac{d\psi}{dt} = \left(\frac{-\hbar^2\nabla^2}{2m} + V_r(t)+ \frac{m\omega^2 (x^2+y^2)}{2}+g|\psi|^2\right)\psi
\end{equation}
Here $\nabla^2$ is the two dimensional laplacian and $g$ denotes the interaction strength. We approximate the interactions as a contact potential, with $g=4\pi \hbar^2 a_s/(m d_\perp)$.  Here $a_s$ is the s-wave scattering length and $d_\perp \sim \sqrt{\hbar/m\omega_z}$ is the size of the wavefunction in the z-direction \cite{cooperreview}.


We can transform into the rotating frame by taking $\psi \to  R(t) \psi$ where $R(t)=e^{-i\phi(t) L_z /\hbar}$ and the angular momentum operator along z is given by $L_z = i\hbar( x\partial_y-y\partial_x)$. 
Noting that $R^\dagger(t) X(t) R(t)=x$ and $R^{\dagger}(t) Y(t) R(t)=y$, we see that under this transformation, the 
deformation becomes time-independent and the GPE reads,  
\begin{equation}\label{rotatingframe}
    i\hbar\frac{d\psi}{dt} = \left(\frac{-\hbar^2\nabla^2}{2m} + V_{\rm eff}(x,y)-\Omega(t) L_z+g|\psi|^2\right)\psi
\end{equation}
where, 
\begin{equation}
    V_{\rm eff}(x,y) =  \frac{m\omega^2 (x^2+y^2)}{2}- \varepsilon m\omega^2x y.
    \label{potential}
\end{equation}
We further adimensionalize by scaling all length scales by $\sqrt{\hbar/(2m\omega)}$ and time by $1/\omega$. We also scale $\psi$ by $1/\sqrt{N}$ such that $\int d^2r_\perp |\psi|^2 = 1$, where $N$ is the total number of particles. The resulting adimensionalized equation is 
\begin{equation}
i\frac{d\psi}{dt} = \left(-\nabla^2 + \frac{x^2}{4}+\frac{y^2}{4}-\frac{\varepsilon xy}{2}- r(t)L_z
 +\tilde g N|\psi|^2\right)\psi
\label{adimgpe}
\end{equation}
Here $L_z = i (x\partial_y-y\partial_x)$,  $r(t)=\Omega(t)/\omega$ and $\tilde g = 8\pi a_s/d_\perp$. From this point, all equations would be expressed in terms of these dimensionless space and time variables. 

The only free parameters in this model are the interaction strength, $\tilde g N$, the strength of the deformation, $\varepsilon$, and the function $r(t)$ which specifies the rotation rate. We choose $\tilde g N = 1000$ and $\varepsilon = 0.125$ in accordance with the experiment in~\cite{fletcher2019geometric}. 
We have studied other values of the 
interaction strength, and find that it does not substantially change the dynamics other than determining the initial size of the condensate.    
The rotation rate, $r(t)$ is ramped up from 0 to 1 in a time $\tau_r$. For $t<\tau_r$ we take  $r(t) = \sin{\frac{\pi t}{2\tau_r}}$, while for longer times we take $r(t)=1$.

Note that in the non-interacting limit ($\tilde g N = 0$), if $\varepsilon\to0$, the GPE can be written as,
\begin{equation}\label{Bfieldhamiltonian}
    i\frac{d\psi}{dt} = \left((i\Vec{\nabla} -q\vec{A})^2 + \frac{(1-r(t)^2)(x^2+y^2)}{4}\right)\psi
\end{equation}
This is reminiscent of the Schrodinger equation for a particle with charge $q$ in the presence of an external vector potential $\vec{A}$ and a residual harmonic trap \cite{fockdarwin,fockdarwin1}. Here $q\vec{A} = r(t)(-y/2,x/2)$ is the effective vector potential in the symmetric gauge. When $r(t)\to1$, we recover the dynamics of a charged particle purely in the presence of a magnetic field. As discussed in Sec.~\ref{intro}, this is the case when the centrifugal force vanishes and the coriolis force mimics the Lorentz force. The eigenstates are Landau level wavefunctions where the particles undergo cyclotron orbits. Each Landau level has a massive degeneracy due to the different possible in-plane center of mass positions. 

\section{Variational Wavefunction Ansatz}\label{wavefnansatz}

We use a time dependent variational wavefunction approach to approximate the solution of Eq.~(\ref{adimgpe}). At all times we use a normalized gaussian ansatz for our wavefunction,  
\begin{equation}
\begin{split}
    \psi = \frac{1}{\sqrt{\pi l_x(t) l_y(t)}}\exp\left(-\frac{(x-\alpha(t)y)^2}{2l_x(t)^2}-\frac{y^2}{2l_y(t)^2}\right) \times\\
    \exp(i(\phi_x(t)x^2 + \phi_y(t)y^2 + \phi_{xy}(t)xy)).
\end{split}
\label{ansatz}
\end{equation}
The dynamics are encoded in six time dependent parameters: $l_x(t),l_y(t)$ represent the size of the condensate while $\phi_x(t),\phi_y(t)$ quantify particle currents in the x and y directions respectively. The $\phi_{xy}(t)$ term represents a quadrupolar flow.  In particular, it serves as a vortex-free irrotational approximation to solid body rotation of the gas~\cite{fetter}. The orientation of the principle axes of the cloud is determined by $\alpha$.  This form is convenient because during the dynamics, the principle axes of the cloud are almost aligned with $x$ and $y$ axes.  In that limit, $\alpha$ corresponds to the 
 slight tilt 
 from the equipotentials of the deformation term in Eq.~\ref{adimgpe}. 
 
 Gaussian ansatzes of this form have been used to obtain dynamical solutions of the Gross-Pitaevskii equation~\cite{ansatz1,ansatz2}.  It should be a good model as long as there are no dynamical instabilites to vortex formation~\cite{stability}.  We compared our variational calculations to numerical integration of the GPE using finite differences, and find good agreement.  We note that, with a fixed grid, the finite difference approach is limited in the length of time it can describe:  Eventually the cloud becomes larger than the simulation volume.  Furthermore the finite difference approach is numerically expensive compared to the variational approach.

\section{Variational Wavefunction dynamics}\label{wavefndynamics}

According to the standard formulation of the time dependent variational principle~\cite{ansatz1}, we calculate the action 
\begin{equation}\label{action}
    S = \int \!dt \int \!dx\,dy\, \left(i\psi^*\frac{d\psi}{dt} - \psi^* H(t)\psi \right)
\end{equation}
as a functional of the parameters $l_x(t),l_y(t),\alpha(t),\phi_x(t),\phi_y(t),$ and $\phi_{xy}(t)$.  We take
\begin{equation}
    H(t) = -\nabla^2 + \frac{x^2}{4}+\frac{y^2}{4}-\frac{\varepsilon xy}{2}-r(t)L_z+\frac{1}{2}\tilde g N|\psi|^2.
\end{equation}
Note the factor of 2 in the interaction term compared to the right hand side of Eq.~(\ref{adimgpe}).
If we extremize Eq.~\ref{action} with respect to $\psi^*$, we recover the Gross-Pittaevskii equation. As described in Appendix~\ref{eoms}, our variational equations of motion are found by making Eq.~(\ref{action}) stationary with respect to the variational parameters.
We obtain six coupled ordinary differential equations corresponding to the six parameters.

\subsection{Initial Conditions}
For our initial conditions, we find the variational ground state in the stationary trap.
We set the rotation rate to zero, $r(0)=0$, and minimize 
\begin{equation}
E= \int dx\,dy   \,\langle H(0)\rangle= \int dx\,dy   \,\psi^* H(0)\psi 
\end{equation}
under the constraint that the total number of particles is fixed.  For the bulk of our results, we use the experimentally relevant values $\tilde g N=1000$ and $\epsilon=0.125$, in which case
$l_x(0) = 5.02$, $l_y(0) = 5.06$, $\alpha(0) = 0.125$, $\phi_x(0) = \phi_y(0) = \phi_{xy}(0) = 0$.
Note, the cloud is almost cylindrically symmetric, with $l_x$ and $l_y$ differing by less than a percent.

\section{Dynamics}\label{dynamicssection}

We use the NDSolve package in Mathematica to numerically  evolve the equations of motion from Appendix~\ref{eoms}. This uses a robust general purpose algorithm with dynamically adjusted step size.  In the experiment, the anisotropy $\epsilon$ is time-independent, and in section~\ref{constanteps} we analyze that case.  The anisotropic term mixes Landau levels, and we can only drive the system into the traditional Lowest Landau level by slowly turning off $\epsilon$ after ramping up $r(t)$.  Thus in section~\ref{varyingeps}, we add this ramp-down step.



\subsection{Constant $\varepsilon$}\label{constanteps}

\begin{figure*}
\includegraphics[height=4.3cm,width=18cm]{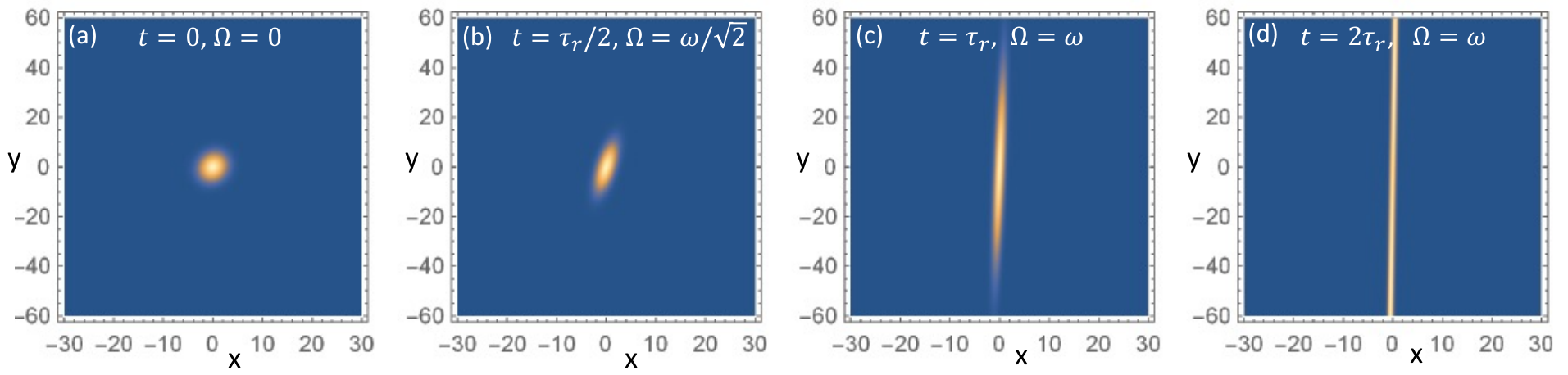}
\caption{Condensate density ($\rho=|\psi|^2$) in the x-y plane as a function of time, 
for a Bose gas in a 2D harmonic potential of frequencty $\omega$ in the presence of a rotating quadrupolar deformation of fixed amplitude $\varepsilon$, as defined in Eq.~(\ref{potential}). As explained after Eq.~(\ref{potential}), $x$ and $y$ are dimensionless, corresponding to using $\sqrt{\hbar/(2m\omega)}$ as the unit length. 
Brighter colors represent higher normalized density $\rho/\rho_{\rm max}$.
From left to right, the rotation rate is ramped up. (a) $\Omega=0$ at time $t=0$ (b) $\Omega=\omega/\sqrt{2}$ at time $t=\tau_r/2$ (c) $\Omega=\omega$ at time $t=\tau_r$ (d) $\Omega=\omega$ at time $t=2\tau_r$. The rotation frequency is held fixed for times longer than $\tau_r$. Here $\varepsilon=0.125$.  
}
\label{densityprofile}
\end{figure*}

Figure~\ref{densityprofile} shows the spatial evolution of the condensate density ($|\psi|^2$) as a function of time for the experimental parameters.
 The condensate expands along the y-direction while getting squeezed in the x-direction.
 
\begin{figure}
\includegraphics[height=12cm,width=7cm]{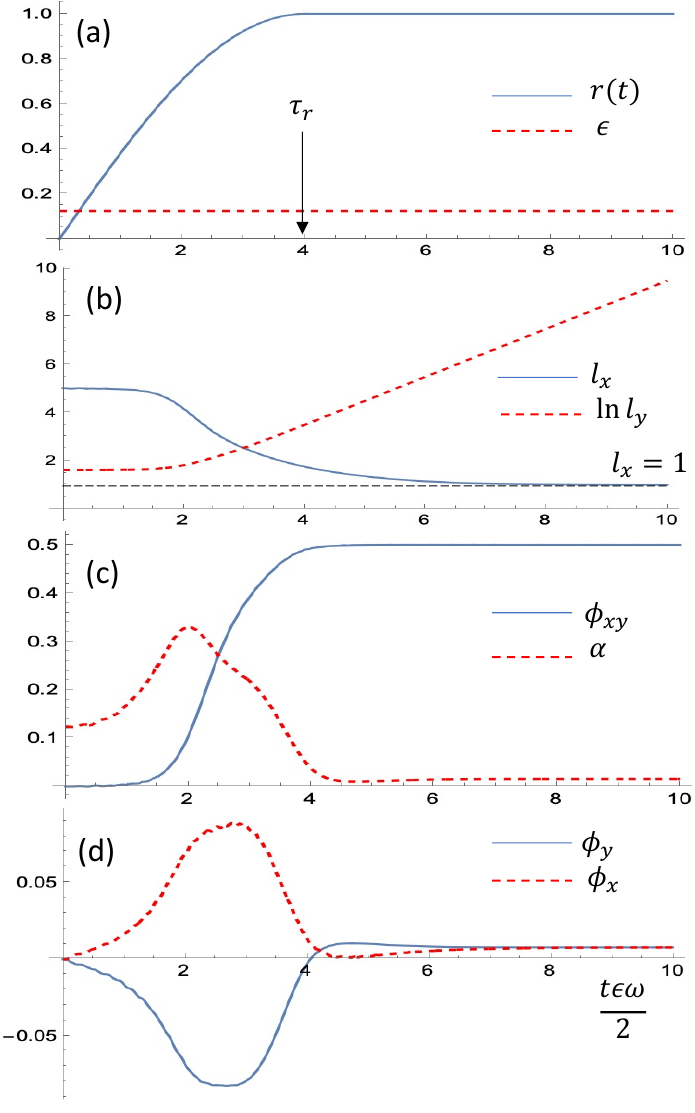}
\caption{Model and Wavefunction ansatz parameters as a function of time for constant $\varepsilon$. All parameters are dimensionless. (a) $r(t)$ (blue,solid) and $\varepsilon = 0.125$ (red,dashed) (b) $l_x(t)$ (blue,solid) saturates at 1 while $l_y(t)$ (red,dashed) grows exponentially at long times (c) $\phi_{xy}(t)$ (blue,solid) saturates at 0.5 and $\alpha(t)$ (red,dashed) saturates at $\varepsilon/8$ (d) $\phi_y(t)$ (blue,solid) and $\phi_x(t)$ (red,dashed) saturate at $\varepsilon/16$ at long times.}
\label{parameters}
\end{figure}

As illustrated in Fig.~\ref{parameters}(a), 
 the dynamics has two regimes: $t<\tau_r$, where the rotation rate $r(t)$ is being swept, and $t>\tau_r$ where the rotation rate is being held constant.  The behavior in the latter regime is particularly simple:  $l_x$ falls exponentially towards 1, $l_y$ grows exponentially,  and $\phi_x,\phi_y,\phi_{xy},$ and $\alpha$ all approach constants.

 This behavior can be understood in terms of an analog of the Lorentz $\vec{E}\times \vec{B}$ drift 
of charged particles in electric and magnetic fields.  In this analogy,
$q\vec{E}= \vec{\nabla}(\varepsilon xy/2)$, and $\vec{B}=r(t) \hat{z} = \hat{z}$ at long times. The $\vec{E}\times \vec{B}$ drift velocities in the x and y direction are respectively given by, $v_x \sim -\varepsilon x/2$ and $v_y \sim \varepsilon y/2$. These equations are consistent  with the exponential squeezing along the x-axis and exponential expansion along the y axis.  This classical argument misses the fact that $l_x$ saturates at 1.  This feature comes from wave mechanics:
The decreasing x-width costs kinetic energy due to the $1/(2l_x^2)$ term and thus there is a {\em quantum pressure} which halts the squeezing.
An increasing y-width doesn't cost any energy in the rotating frame due to the cancellation of the external harmonic trap with the centrifugal force. At long times, as $l_y \rightarrow \infty$, the kinetic energy, $1/(2l_y^2)$ and interaction energy $\tilde g N/4\pi l_x l_y$ tend to zero. 

At long times, $\alpha(t) \rightarrow \varepsilon/8$ denoting a finite tilt of the condensate away from the y axis. This wavefunction tilt can be understood from a slightly more elaborate analysis of the classical equations of motion.  
For this model, the Lorentz force equation is linear,
$m d\vec{v}/dt = q\vec{E} + q\vec{v}\times\vec{B}$, and Fourier transforming it in the time domain
yields
\begin{equation}\label{freqeqn}
    -\nu^2\begin{bmatrix} \Bar{x}(\nu)\\ \Bar{y}(\nu) \end{bmatrix} = \begin{bmatrix} 0 & \varepsilon + 2 i \nu\\ \varepsilon - 2i \nu & 0 \end{bmatrix} \begin{bmatrix} \Bar{x}(\nu)\\ \Bar{y}(\nu) \end{bmatrix}
\end{equation}
with 
$\Bar{x}(\nu)=\int dt\, e^{i \nu t} x(t)$, and a similar expression for $\bar y$.
From Eq.~(\ref{freqeqn}), we find a quartic equation for the allowed frequencies of free oscillation, $\nu$.  As $\epsilon\to 0$, the solutions are 
$\nu=\pm i \varepsilon/2,\pm 2$.  The first solutions are purely imaginary and represent the squeezing already discussed.  The latter describe cylclotron motion.  The eigenvector of the exponentially growing solution has $\bar y/\bar x=\varepsilon/8$, corresponding to the observed tilt.  This is analogous to inertial drift in plasma physics~\cite{plasma}.


The modulus and phase of the condensate wavefunction can be interpreted in terms of fluid mechanics, $\psi =\sqrt{\rho}e^{i\Phi}$, where $\rho$ is the density and ${\vec v} =(h/m)\vec\nabla \Phi$.    
Hence, in dimensionless units, our ansatz corresponds to a  flow-pattern $v_x= 2(2\phi_x x+\phi_{xy} y)$ and
$v_y= 2(2\phi_y y + \phi_{xy} x)$.
When $\phi_{xy}=0.5$, the angular momentum of the condensate resembles that of solid body rotation of the condensate (see  Appendix~\ref{solidbodyrotation}). In Fig.~\ref{parameters}(d), one sees that the particle current terms $\phi_y$ and $\phi_x$ saturate at $\varepsilon/16$ at long times. 

Further insight into the structure of the wavefunction at long times comes from transforming into a coordinate system which is aligned with the principle axes of the condensate.
We define  $\tilde x = (x-\alpha y)$ and $\tilde y = (y+\alpha x)$, and take the parameters to have their long-time values,
 $\phi_x=\phi_y=\varepsilon/16$, $\alpha=\varepsilon/8$, and $\phi_{xy}=1/2$, $l_x=1$.
Up to linear order in $\varepsilon$, the variational wavefunction is then
\begin{equation}\label{longtimewavefn}
    \psi \sim \frac{1}{\sqrt{\pi \ell_y}}\exp\left(-\frac{\tilde x^2}{2}+i\frac{\varepsilon \tilde y^2}{8}+i\frac{\tilde x \tilde y}{2}-\frac{\tilde y^2}{2 l_y^2}\right)
\end{equation}
At long times $l_y\to\infty$, and one can ignore the last term in the exponential.
The three remaining terms have physical meanings.  The first limits the spread of the particles in the $\tilde x$ axis.
The second represents out-going currents in the $\tilde y$ axis, corresponding to the expansion. The third is an irrotational flow pattern that approximates solid body rotation. If $\epsilon=0$ and $l_y \to \infty$, this is in the lowest Landau level, in the symmetric gauge.  Transforming to the Landau gauge corresponds to multiplying by $\exp(-i\tilde x\tilde y/2)$, which eliminates the third term.  

In our 'squeezing' solution of the classical Lorentz force equations earlier in the section, the rate of exponential expansion was $\varepsilon/2$. This is consistent with $\phi_y=\varepsilon/16$ at long times since the outgoing velocity in the $\tilde y$ direction becomes $\varepsilon \tilde y/2$ by calculating gradient of the phase in Eq.~(\ref{longtimewavefn}). Due to a finite wavefunction tilt and exponential expansion, we expect $\phi_x = \phi_y$ at long times.

\begin{figure}
\includegraphics[width = \columnwidth]{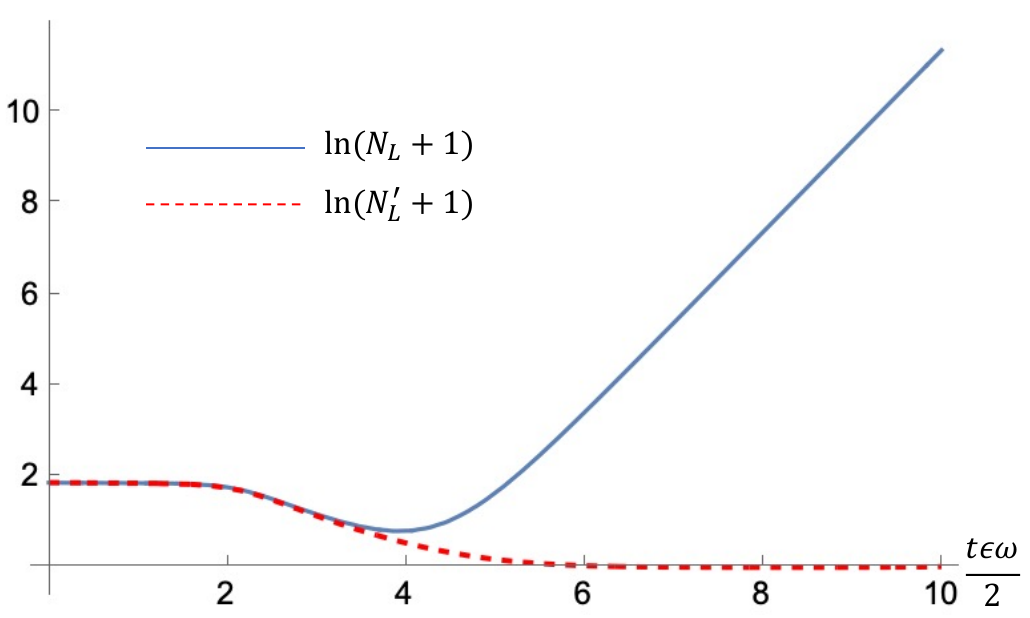}
\caption{Logarithm of the average traditional Landau level occupation, ln$(N_L+1)$ (blue,solid) and logarithm of the average distorted Landau level occupation, ln$(N_L^\prime+1)$ (red,dashed), as a function of time for the ramp described in Fig.~\ref{parameters}(a). At long times, $N_L$ grows exponentially while $N_L^\prime$ decreases and tends to 0.}
\label{landaulevels}
\end{figure}
The saturation of $l_x$ is consistent with being in the Lowest Landau level.  There are some subtleties however, as the eigenstates of the deformed trap are somewhat different than those of traditional Landau levels.  We can quantify this feature by
calculating the average number of landau levels which are occupied,
\begin{equation}\label{N_L}
    N_L = \int dx\,dy   \,\psi^*\frac{(H_{L}-1)}{2}\psi
\end{equation}
where
\begin{equation}
     H_{L} =  -\nabla^2 + \frac{x^2}{4}+\frac{y^2}{4}-i(x\partial_y - y\partial_x)
\end{equation}
is the symmetric gauge Hamiltonian of a charged particle in a uniform magnetic field: $N_{L} = 0$ when the wavefunction is in the lowest Landau level. We plot ln$(N_L + 1)$ as a function of time in Fig.~\ref{landaulevels}.  As the rotation rate increases, $N_L$ drops.  However, after the ramp is concluded ($t\varepsilon\omega/2=4$), the average Landau level number grows exponentially.  This behavior is evident from Eq.~(\ref{longtimewavefn}), for which $N_L\sim\varepsilon^2 l_y^2/16$.  Since $l_y$ grows exponentially, so does $N_L$.  

Following the supplementary of~\cite{fletcher2019geometric}, one can alternatively use a gauge transformation to define distorted Landau levels which take into account the elliptic deformation of the trap. The gauge transformation is given by $\psi^\prime = G\psi$ where $G = \exp(i\varepsilon(x^2-y^2)/8)$. One can then define the distorted Landau levels and count their average occupation via
\begin{equation}
N_L^\prime=\int dx\,dy   \,\psi^{\prime*}\frac{(H_{L}^\prime-1)}{2}\psi^\prime
\end{equation}
where
\begin{equation}
     H_{L}^\prime =  -\frac{\nabla^2}{\gamma} + \gamma\frac{(x^2+y^2)}{4}-i(x\partial_y - y\partial_x),
\end{equation}
with $\gamma = \sqrt{1+\varepsilon^2/4}$. We plot ln$(N_L^\prime + 1)$ as a function of time in Fig.~\ref{landaulevels}. In terms of these  distorted Landau levels, $N_L^\prime$ drops as the rotation rate increases and tends to 0 at long times. The condensate is thus driven into this distorted lowest Landau level.  

When $l_y\to\infty$,
the difference between Eq.~(\ref{longtimewavefn}) and an undistorted lowest Landau level wavefunction  is the phase factor $\exp(i\varepsilon \tilde y^2/8 )$, which, as already explained, quantifies the outward expansion of the condensate.  This expansion is driven by the Lorentz $E\times B$ drift, which occurs as long as $\varepsilon\neq 0$.  Thus if we wish to  end up in the undistorted lowest Landau level, we must ramp $\varepsilon$ to zero.  In the next section we include such a ramp, and analyze the resulting dynamics.


\subsection{Varying $\varepsilon$}\label{varyingeps}

We adiabatically turn off the anisotropy $\varepsilon$ after the rotation rate $r(t)$ has been ramped up to 1. For $t<\tau_r$, $\varepsilon(t) = \varepsilon_0 = 0.125$. For $\tau_r < t < \tau_r + \tau_\varepsilon$, $\varepsilon(t) = \varepsilon_0\frac{\tau_r+\tau_\varepsilon-t}{\tau_\varepsilon}$. It is held at zero for all later times. Thus $\varepsilon(t)$ reaches zero at time $t = \tau_r + \tau_\varepsilon$. 

\begin{figure}
\includegraphics[width=0.48\textwidth]{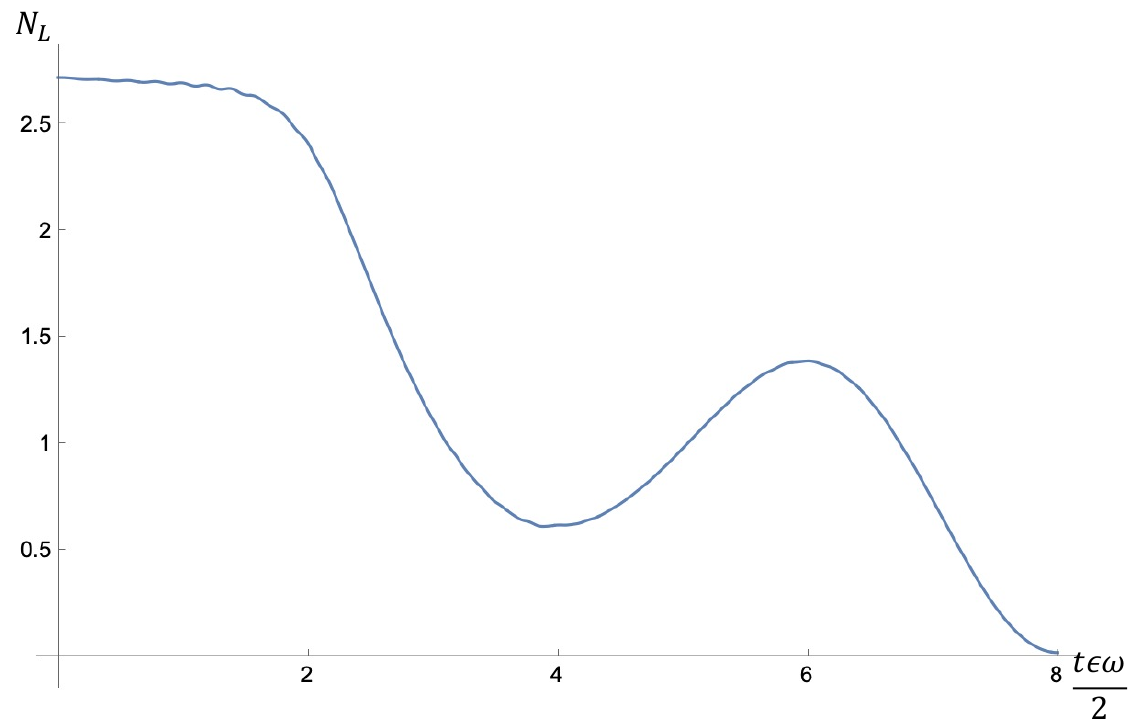}
\caption{Average Landau level occupation, $N_L$, as a function of time, when the trap asymmetry is ramped to zero.  Parameters are shown in Fig.~\ref{parametersvary}. $N_L = 0.03$ at the end, implying particles having high probability of being in the lowest Landau level.}
\label{landaulevelsvary}
\end{figure}

\begin{figure}
\includegraphics[height=12cm,width=7cm]{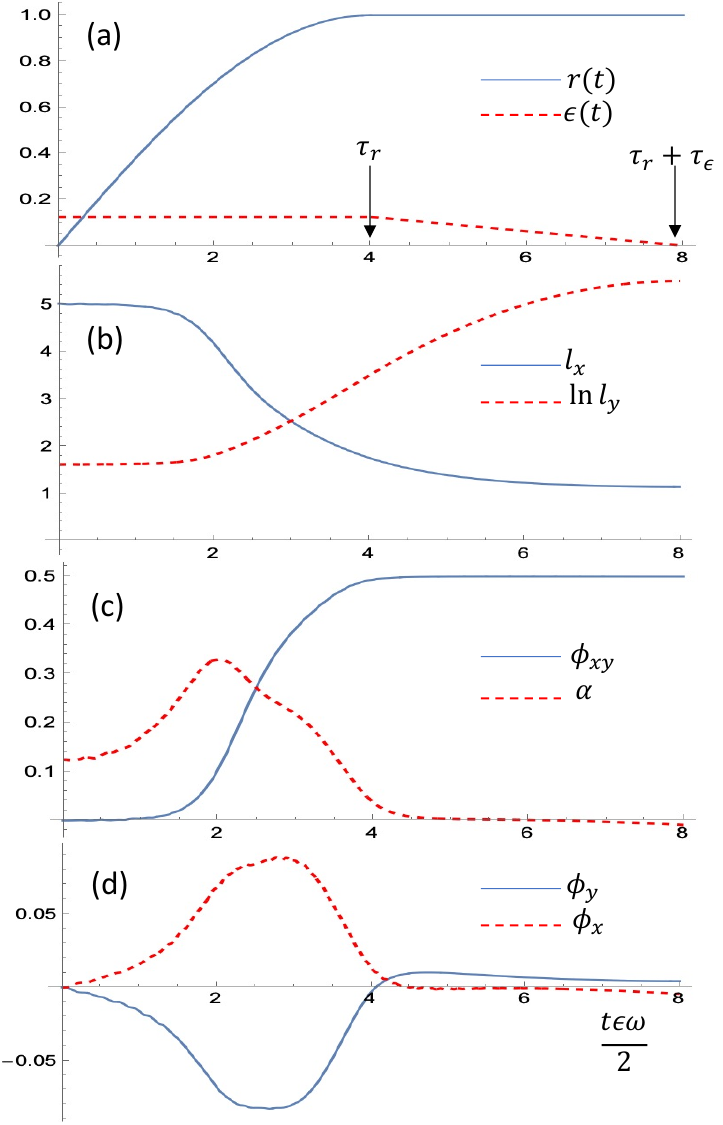}
\caption{Model and Wavefunction ansatz parameters as a function of time in the protocol where $\varepsilon$ is ramped to zero. All parameters are dimensionless. (a) $r(t)$ (blue,solid) and $\varepsilon$ (red,dashed). (b) $l_x(t)$ (blue,solid) and ln$(l_y(t))$ (red,dashed); ramping $\varepsilon\to 0$ cuts off the exponential growth of $l_y$ that was seen in Fig.~\ref{parameters}. (c) $\phi_{xy}(t)$ (blue,solid) again saturates at 0.5 and $\alpha(t)$ (red,dashed) saturates at a negative value which depends on the ramp. (d) $\phi_y(t)$ (blue,solid) and $\phi_x(t)$ (red,dashed) saturate at $\pm \alpha/2$.  All parameters are subsequently time independent.}
\label{parametersvary}
\end{figure}

For adiabaticity, we need $\tau_\varepsilon > 1/\omega$. In order to enter the lowest Landau level, we want the squeezing dynamics to continue until $l_y \gg 1$ and $l_x \sim 1$. Thus $\tau_\varepsilon$ needs to be moderately larger than $1/(\varepsilon_0\omega)$. 
In Fig.~\ref{landaulevelsvary} and \ref{parametersvary}, we consider the case where $\tau_\epsilon=8/(\varepsilon_0\omega)$.  We only show data  up to $t=\tau_r+\tau_\varepsilon$ as all parameters are time independent once $\varepsilon=0$.  Remarkably, as shown in Fig.~\ref{landaulevelsvary}, this ramp results in a final $N_L\approx 0.03$, which means that the probability of a particle being in the lowest Landau level exceeds 97\%.  

For these parameters, $N_L$ in Fig.~\ref{landaulevelsvary} is a non-monotonic function of time.  The exponential growth,  seen in  Fig.~\ref{landaulevels} competes with the expected drop due to reducing $\varepsilon$.  As described below, this competition also results in a non-monotonic dependence of the final value of $N_L$ on  $\tau_\epsilon$.

The system and wavefunction parameters as a function of time are shown in Fig.~\ref{parametersvary}. Similar to the constant $\varepsilon$ case in Fig.~\ref{parameters}(b), $l_x$ falls exponentially towards 1, and $l_y$ grows.  For moderate times, the growth is roughly exponential, but it flattens out and saturates as $t\to\tau_r+\tau_\varepsilon$.  As before, $\phi_{xy} \to 0.5$. The wavefunction tilt, parameterized by $\alpha$, switches directions during the $\varepsilon$ ramp.  Unlike the findings in Fig.~\ref{parameters}(c), the final tilt depends on details of the ramp.

The particle current terms, $\phi_x$ and $\phi_y$ take values $-\alpha/2$ and $\alpha/2$ respectively. 
Transforming into the tilted coordinates again (treating $\alpha$ as small) gives the wavefunction in Eq.~(\ref{longtimewavefn}), but this time  with $\varepsilon=0$.  This is a lowest Landau level wavefunction.  There is, however, a correction term of higher order in $\alpha$ in the wavefunction.  It is given by $\exp(i\alpha^3 (\tilde x^2-\tilde y^2)/2)$. Due to this correction, the leading order correction to $N_L$ from Eq.~(\ref{N_L}) is given by $\alpha^4 l_y^2$. Thus as long as $\alpha^4 l_y^2 \ll 1$, $N_L \sim 0$. Since $\alpha \sim \varepsilon_0/8$ and $l_y \sim e^{\varepsilon_0\omega t/2}$, this condition limits the length of the ramp down period of $\varepsilon$ to about $8/(\varepsilon_0 \omega)\ln(8/\varepsilon_0)$. When $\varepsilon$ is switched off on this timescale, the condensate is left in the undistorted lowest Landau level at the end.

\section{Summary and Outlook}

We modeled the dynamics of a condensed Bose gas in a rotating anisotropic trap. We used a time dependent variational wavefunction 
to calculate the time evolution of the system. We find that the behavior at long times is well described by the classical equations of motion of a charged particle in crossed electric and magnetic fields.  We calculate the average Landau level occupation as a function of time. We find that when the trap anisotropy is constant in time, the condensate gets squeezed into a long thin shape which occupies a deformed lowest Landau level.
If the trap anisotropy is switched off at the appropriate time, the condensate enters the traditional lowest Landau level. 

Confinement to the Lowest Landau level is a necessary prerequisite for studying analogs of quantum Hall physics with cold atoms.  We emphasize, however, that the states produced by this protocol are rather different from the homogeneous liquids which lead to the quantum Hall effect.  Nonetheless, this is an important first step. Moreover there are a number of interesting purely interactions-driven phenomena which can be explored starting from these highly anisotropic lowest Landau level states.  For example, the longer-time dynamics are described by correlated hopping models with unique sets of conserved quantities that lead to unusual transport~\cite{correlatedhopping}.  Variants of this protocol can also lead to interesting pattern formation~\cite{mukherjee2021crystallization}.
%

\section{Acknowledgements}

We thank Martin Zwierlein and his research group for insightful discussions.
This work was supported by the NSF Grant No. PHY- 2110250.

\appendix

\section{Variational equations of motion}\label{eoms}

Using time dependent variational principle, we first calculate the action given by, 
\begin{eqnarray}
    S &=& \int \!\!dt\! \int \!\!dx\,dy\, \left(i\psi^*\frac{d\psi}{dt} - \psi^* H(t)\psi \right)\\
    &=&
    \int\!\!dt\, \mathcal{F}(l_x(t),l_y(t),\alpha(t),\phi_x(t),\phi_y(t),\phi_{xy}(t))
    \nonumber
\end{eqnarray}
where the variational wavefunction ansatz $\psi$ is given by Eq.~(\ref{ansatz}), 
\begin{equation}\label{ansatzappendix}
\begin{split}
    \psi = \frac{1}{\sqrt{\pi l_x(t) l_y(t)}}\exp\left(-\frac{(x-\alpha(t)y)^2}{2l_x(t)^2}-\frac{y^2}{2l_y(t)^2}\right) \times\\
    \exp(i(\phi_x(t)x^2 + \phi_y(t)y^2 + \phi_{xy}(t)xy))
\end{split}
\end{equation}
The spatial integrals in the action are gaussian and can be done analytically, yielding
$\mathcal{F}=\mathcal{F}_t+\mathcal{F}_p+\mathcal{F}_{\rm trap}+\mathcal{F}_{\rm rot}+\mathcal{F}_{\rm int}$,
\begin{align*}
    \mathcal{F}_t =& -\frac{d\phi_x}{dt}\left(\frac{l_x^2}{2}+\frac{\alpha^2 l_y^2}{2}\right) - \frac{d\phi_y}{dt}\frac{l_y^2}{2} - \frac{d\phi_{xy}}{dt} \frac{\alpha l_y^2}{2} \\
    \mathcal{F}_p =&
    \frac{1+\alpha^2}{2l_x^2}+\frac{1}{2l_y^2}+ 
    l_x^2\left(
    2\phi_x^2+\frac{\phi_{xy}^2}{2}
    \right)\\
    + l_y^2 &\left(
    2\alpha^2 \phi_x^2
    +2\phi_y^2
    +2\alpha\phi_{xy}(\phi_x+\phi_y)+
    \phi_{xy}^2 \frac{1 
    + \alpha^2}{2}
    \right)
\\    
\mathcal{F}_{\rm rot}=&r\left(\alpha(\phi_x+\phi_y)l_y^2+\phi_{xy}\frac{l_x^2+(\alpha^2-1)l_y^2}{2}\right)\\
\mathcal{F}_{\rm trap}=&
     -\varepsilon\alpha l_y^2/4+\frac{l_x^2}{8}+\frac{\alpha^2l_y^2}{8}+\frac{l_y^2}{8}\\
   \mathcal{F}_{\rm int} =& \frac{\tilde g N}{2\pi l_xl_y}
\end{align*}

We extremize the action using the  Euler-Lagrange equations,
\begin{equation}
    \frac{\partial \mathcal{F}}{\partial \Phi} - \frac{d}{dt}\frac{\partial \mathcal{F}}{\partial \Dot{\Phi}}=0    
\end{equation}
where $\Phi = \{l_x,l_y,\alpha,\phi_x,\phi_y,\phi_{xy}\}$ and $\Dot{\Phi}$ denotes a time derivative. We thus get six coupled nonlinear ordinary differential equations. We numerically solve these equations.

\section{Parameter $\phi_{xy}$ approximating solid body rotation}\label{solidbodyrotation}
The term $\exp(i \phi_{xy} x y)$ in Eq.~(\ref{ansatzappendix}) represents an irrotational quadrupolar flow. This term contributes to a non-zero angular momentum, $L_z^c$ given by, 
\begin{equation}
    L_z^c = \langle\psi|L_z|\psi\rangle = \frac{\phi_{xy}(l_y^2 - l_x^2)}{2}
\end{equation}
If the condensate were rotating as a solid body with angular velocity $\Omega$, condensate velocity profile is $\vec{v} = \vec{r} \times \vec{\Omega}$ and the solid body angular momentum $L_z^{sb}$ is $m r^2 \Omega$. In dimensionless units, $L_z^{sb}$ is,
\begin{equation}
    L_z^{sb} = \langle \psi | \frac{r^2}{2}|\psi \rangle = \frac{(l_y^2 + l_x^2)}{4} 
\end{equation}

The condensate approximates solid body rotation when $l_y \gg l_x$ and $\phi_{xy}=0.5$, making $L_z^c/ L_z^{sb} \to 1$.

\bibliography{main.bib}

\end{document}